\documentclass[a4paper,fleqn,usenatbib]{mnras}

\usepackage{newtxtext,newtxmath}

\usepackage[T1]{fontenc}
\usepackage{ae,aecompl}

\usepackage{graphicx}	
\usepackage{amsmath}	
\usepackage{amssymb}	
\usepackage{multicol} 
\usepackage{pdflscape}

\title[Pushy neighbors]
{Companion-driven evolution of massive stellar binaries }

\author[Rose et al.]{
Sanaea C.~Rose,$^{1}$\thanks{E-mail: srose@astro.ucla.edu}
Smadar Naoz,$^{1,2}$ and
Aaron M.~Geller$^{3,4}$
\\
$^{1}$Department of Physics and Astronomy, University of California, Los Angeles, CA 90095, USA\\
$^{2}$Mani L. Bhaumik Institute for Theoretical Physics, Department of Physics and Astronomy, UCLA, Los Angeles, CA 90095, USA\\
$^{3}$Center for Interdisciplinary Exploration and Research in Astrophysics (CIERA) and Department of Physics and Astronomy, \\ Northwestern University, 2145 Sheridan Road, Evanston, IL 60201, USA \\
$^{4}$Adler Planetarium, Department of Astronomy, 1300 S. Lake Shore Drive, Chicago, IL 60605, USA \\
}

\date{Accepted XXX. Received YYY; in original form ZZZ}

\begin{document}
\label{firstpage}
\pagerange{\pageref{firstpage}--\pageref{lastpage}}
\maketitle

\begin{abstract}

At least $70\%$ of massive OBA-type stars reside in binary or higher-order systems. The dynamical evolution of these systems can lend insight into the origins of extreme phenomena such as X-ray binaries and gravitational wave sources. In one such dynamical process, the Eccentric Kozai-Lidov Mechanism, a third companion star alters the secular evolution of a binary system. For dynamical stability, these triple systems must have a hierarchical configuration. We explore the effects of a distant third companion's gravitational perturbations on a massive binary's orbital configuration before significant stellar evolution has taken place ($\leq 10$~Myr). We include tidal dissipation and general relativistic precession. With large ($38,000$ total) Monte-Carlo realizations of massive hierarchical triples, we characterize imprints of the birth conditions on the final orbital distributions. Specifically, we find that the final eccentricity distribution over the range $0.1-0.7$ is an excellent indicator of its birth distribution.  Furthermore, we find that the period distributions have a similar mapping for wide orbits. Finally, we demonstrate that the observed period distribution for approximately $10$~Myr-old massive stars is consistent with EKL evolution.
\end{abstract}


\section{Introduction}

Recent observations suggest that massive stellar binaries are prevalent in our Galaxy. In fact, more than $70\%$ of OBA and $50\%$ of KGF spectral type stars likely exist in binaries \citep[e.g.,][]{Tok08,Raghavan+10,Sana+12}. Observations of massive binaries suggest unique orbital parameter distributions compared to KGF binaries. For example, \citet{DK2013} estimate that $30 \%$ of massive stellar binaries have periods less than ten days, while some power law can represent the slow decline in the number of systems out to about $10^4 \, \mathrm{AU}$. \citet{Sana+12} also find that the period distribution of OBA stars increases dramatically toward smaller periods. These findings contrast with those for KGF stars, which populate a log-normal period distribution peaked around $10^5$ days \citep{Raghavan+10}. While the period distributions for these spectral types differ, the eccentricity distribution of OBA stars may be closer to that of the KGF stars. See Section \ref{Observations} for a more detailed discussion.

Many short period KGF binaries may in fact occur in a triple configuration \citep{T97,Pri+06,Tok08,Moe,Egg+07,Griffin2012}. Similarly, massive binaries may often reside in triple configurations \citep{Raghavan+10,Zinnecker&Yorke}. While the fraction of massive stars in triples is not well known, the average number of companions per OB primary may be at least three times higher than that of low-mass stars, which have $0.5$ companions on average \citep{Preibisch+01,Grellman+13}.

Dynamical stability arguments imply that these triple systems must have a hierarchical configuration: the third body orbits the inner binary on a much wider outer orbit. In this configuration, coherent gravitational perturbations from the outer body influence the long-term evolution of the inner orbit. The orbits can be treated as massive wires that torque each other, where the line-density of each wire is inversely proportional to the orbital velocity. In this orbit-averaged, or secular, approximation, the semi-major axis ratio remains constant. The gravitational potential can be expanded in terms of this ratio $a_1/a_2$, where $a_1$ ($a2$) is the semi-major axis of the inner (outer) orbit \citep{Kozai,Lidov}. The hierarchical configuration makes this expansion possible by ensuring that $a_1/a_2$ is a small parameter. The lowest order, or quadrupole level, of approximation is proportional to $(a_1/a_2)^2$. The next level of approximation is called the octupole level \citep[see][a recent review]{Naoz16}.

The applications of hierarchically configured stellar triples have been explored extensively in the literature \citep{NF,Harrington1969,MazehShaham1979,Kiseleva1998,FabryckyTremaine07,PeretsFabrycky09,Thomson2011,Naoz+13sec,Shappee+13,Pejcha13,MichaelyPerets14,Prodan+15,Stephan+16,MoeKratter18,Bataille+18}. Furthermore, many studies suggest that short period compact object binaries, including black hole, neutron star, and white dwarf binaries, may result from hierarchical triple dynamics \citep{Thomson2011,KatzDong2012,Antonini&Perets12,Naoz+16,Toonen+18,Hoang+18}. We examine the impact of EKL-driven dynamical evolution on the properties of massive binaries embedded in triples. The birth orbital configurations of massive binaries are not well constrained. We therefore explore different possibilities for the period and eccentricity birth distributions with the objective of discerning the birth properties from the final results. We test a broad range of initial distributions, tidal recipes, and efficiencies to identify the manner in which observed period and eccentricity distributions map back onto the initial distributions. We focus on $10$~Myr old systems, i.e., before significant stellar evolution has taken place. Our simulations include tidal dissipation and general relativistic precession.

We run $38$ sets of Monte-Carlo numerical simulations with a variety of initial conditions and tidal recipes. Each simulation has $1000$ systems, bringing the total number of realizations to $38,000$. Section \ref{method} reviews our methods and initial conditions. In Section \ref{Results}, we present our results and analyze the statistical distributions of the inner orbital periods and eccentricities. Section \ref{Signatures} investigates traces of the initial conditions in the final distributions, traces which persist even after $10$~Myr of EKL-driven evolution. Finally, we consider our simulated results in the context of observations and discuss the implications in Section \ref{Observations}. Our simulations suggest that the EKL mechanism can re-distribute the orbital periods to match observed distributions, while the eccentricity distribution retains its original shape.


\section{Methodology, Numerical Setup and Initial Conditions} \label{method}

\subsection{Point mass dynamics}
We solve the secular equations of motion for the hierarchical triple to the octupole-level of approximation, as described in \citet{Naoz+13sec}. Stars with masses $m_1$ and $m_2$ compose the inner binary, while the tertiary body with mass $m_3$ and the inner binary form an outer binary. The inner (outer) orbit has the following parameters: $a_1$ ($a_2$), $e_1$ ($e_2$),  $\omega_1$ ($\omega_2$) and $i_1$ ($i_2$) for the semimajor axis, eccentricity, argument of periapsis, and inclination with respect to the total angular momentum, respectively. We define $i_{tot}$ as $i_1+i_2$. We include general relativistic precession following \citet{Naoz+13GR}, who demonstrate that GR precession can suppress or facilitate eccentricity excitations in different parts of phase space. 

We limit our simulations to $10$ million years to allow for comparisons with observed massive stars in young stellar clusters \citep[e.g.,][]{Sana+12}. We treat the mass and radius of each star as constant. Over ten million years, a $20~M_{\odot}$ star, the largest mass allowed in these simulations, will lose about three percent of its mass, and the semi-major axis of its orbit will expand by three percent. Thus, while the interplay between the point-mass dynamics and stellar evolution has interesting consequences \citep[e.g.,][]{Stephan+16}, it has negligible effect on our calculations. 

\subsection{Tidal models}

We include tidal dissipation, which acts to circularize and tighten the inner binary. 
Following  \citet{NF}, we adopt the tidal evolution equations of \citet{Eggleton2001}. These equations follow the equilibrium (E) tide model of \citet{Egg1998}. \citet{Eggleton2001} relate the viscous timescale $t_v$ of a star to the tidal dissipation timescale $t_F$ using the Love parameter $k_{L,j}$, where $j=1,2$ for masses $m_1$ and $m_2$, respectively. The Love parameter quantifies the quadrupolar deformability of a star. For the primary star with mass $m_1$,
\begin{eqnarray}
t_{F,1} =t_{v,1} \left( \frac{a_1}{R_1} \right)^{8} \frac{m_1^2}{m_2(m_1+m_2)} \frac{1}{ 9(1+2k_{L1})^2 } \ .
\end{eqnarray}
We use $k_{L,j} = 0.014$ for an $n=3$ polytrope \citep{Eggleton2001}. Both the primary and secondary stars have spin periods of four days. Based on \citet{Bouvier}, the spin period of a massive star is approximately two days. A factor of two has minimal impact in these tidal prescriptions: The system in the right-hand panel of Figure \ref{fig:realtime} exhibits the same behavior if we set the stellar spins to two days instead of four.

We also implement a prescription for radiatively-damped dynamical (RDD) tides \citep{Zahn77} for stars of mass greater than $1.5$~M$_{\odot}$. 
Following \citet{Hurley+02} and \citet{Zahn77}, we express the tidal dissipation timescale as:
\begin{eqnarray}
t_F &=&  \left( \frac{a_1}{R_1} \right)^{9} \sqrt{\frac{a_1^3}{Gm_1}} \frac{m_1}{m_2}\left(1+\frac{m_2}{m_1}\right)^{11/6} \nonumber \\ 
&\times& \frac {1} {9 (1.592\times10^{-9})}\left(\frac{ m_1}{{\rm M}_\odot}\right)^{2.84} \ .
\end{eqnarray}
We note that in a comparison with numerical calculations, \citet{Chernov} suggests that this tidal prescription underpredicts the tidal efficiency in short period binaries. \citet{ClaretCunha1997} find that Zahn's formalism fails to account for some observed circularized systems. The treatment of stellar tides has invoked much debate \citep{Langer2009}. Noting this lack of accord, \citet{Yoon+10} include a parameter for the rate of tidal synchronization to test for its effects. Similarly, we opt to use the equilibrium tide model for all stars in select simulations, allowing us to treat the viscous timescale as a free parameter and tune the tidal efficiency. 

We label our simulations according to tidal prescription and efficiency as detailed in Table \ref{tab:labels}. Both \citet{Hurley+02} and \citet{Zahn77} note the inefficient nature of RDD tides compared to E tides. We use E tides with $t_v = 0.005$, an unrealistically small value, in a few simulations for comparative purposes. 

\begin{table}
 \caption{Simulations are named according to tidal prescription and efficiency. In IET and LET simulations, RDD tides are used for stars with masses greater than $1.5 M_{\odot}$, while E tides with the specified $t_v$ are used for lower mass stars.}
 \label{tab:labels}
 \begin{tabular}{ll}
  \hline
  Label: Name & Tidal Prescription\\
  \hline
  LET: Least Efficient Tides & RDD or E $t_v = 50$~yr  \\[0.8pt]
  IET: Inefficient Tides & RDD or E $t_v = 0.5$~yr \\[0.8pt]
  MET: Moderately Efficient Tides & E $t_v = 50$~yr \\[0.8pt]
  SET: Super Efficient Tides & E $t_v = 0.5$~yr \\[0.8pt]
  UET: Unrealistically Efficient Tides & E $t_v = 0.005$~yr  \\[0.8pt]
  \hline
 \end{tabular}
\end{table}

\subsection{Birth Distributions}

To explore the influence of initial parameter distributions on the results, we use several different probability distributions to draw initial conditions for the Monte Carlo simulations. We label simulations according to their birth distributions.

The simulations fix the mass of the primary $m_1$ at $10~M_{\odot}$. We then draw $m_2$ and $m_3$ from uniform distributions of the mass ratios $q_1= m_2/m_1$ and $q_2= m_3/(m_1+m_2)$ where $0.1<q<1$. We discuss simulations that use a Kroupa mass function in Appendix \ref{appendix:table}. A power law of the form $R \propto M^{\alpha}$ derives the stellar radii from the masses. For most simulations, we use the ZAMS mass-radius relation $R = 1.01 M^{0.57}$\footnote{For comparison purposes, below, we also consider larger radii using $R = 1.33 M^{0.55}$, an empirical fit to observations, and $R = 1.61 M^{0.81}$, the TAMS mass-radius relation \citep{Demircan}.}. In all simulations, we include a spin-orbit misalignment for the primary (secondary) star $\beta_{1}$ ($\beta_2$), which we draw from an isotropic distribution. 

\begin{figure}
    \centering
        \includegraphics[width=.48\textwidth]{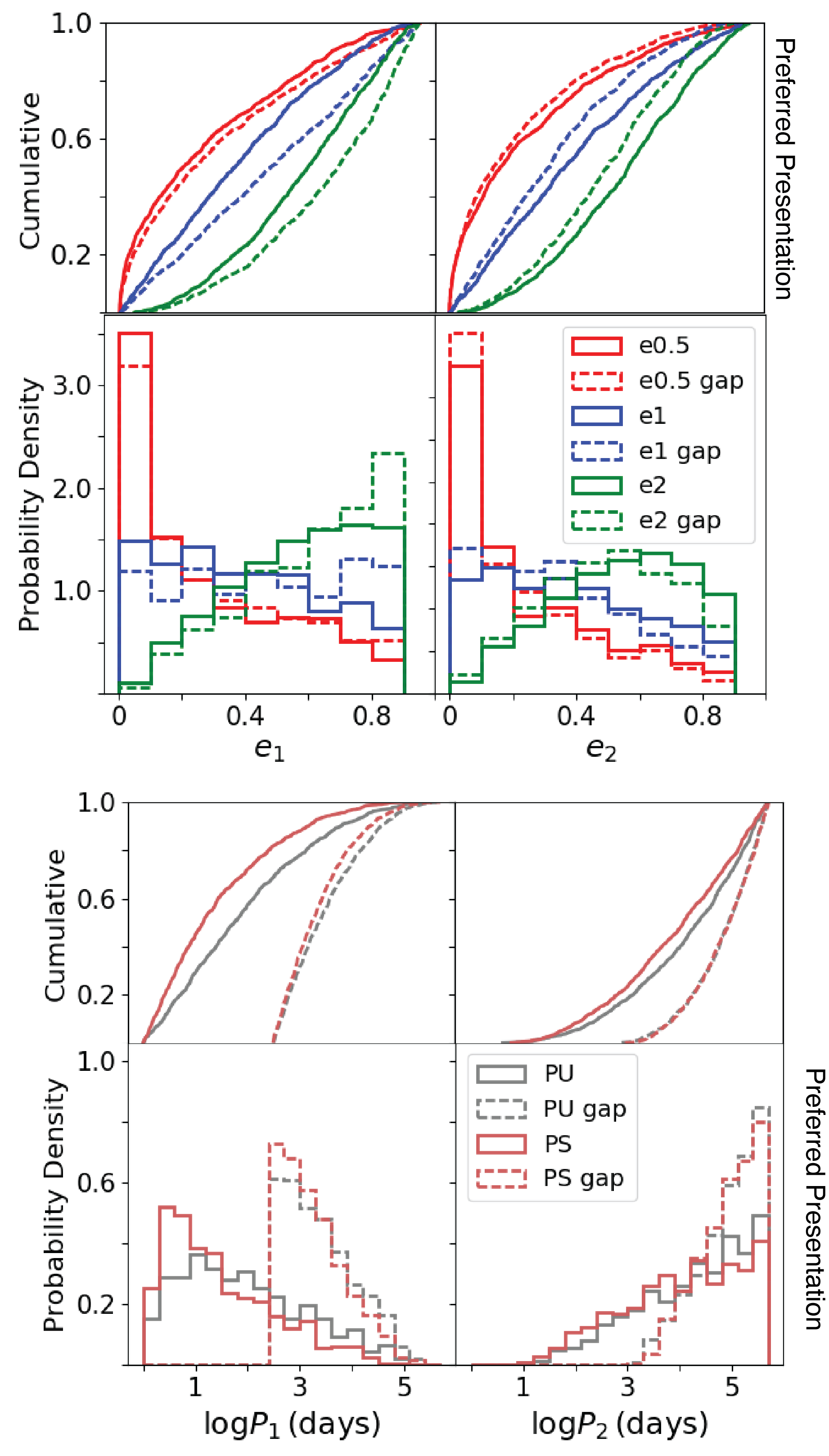}
    ~ 
    \caption{The possible initial conditions for period and eccentricity.
    {\bf Top panel:} Cumulative eccentricity distribution {\bf Second panel:} Eccentricity probability density {\bf Third panel:} Cumulative period distribution {\bf Third panel:} Period probability density. We consider the initial distributions after applying stability criteria (see text). Initial conditions with a gap are represented by dashed lines. In the top two panels, thermal, uniform, and $\sim e^{-0.5}$ are shown in green, blue, and red, respectively. Note that those simulations with no gap have fewer high eccentricity systems, a result of having smaller values of $a_{1i}$ while still needing to satisfy the Roche limit criterion. In the bottom panels, we show a uniform distribution in logarithmic period (grey lines) and $f(\log {P}/{\textrm{days}}) \propto (\log {P}/{\textrm{days}})^{-0.5}$ (crimson lines) \citep{Sana+12,Sana2013}. Since the stability criteria tend to curtail large values of $a_1$, the PU and PS gap distributions appear similar. We observe a similar albeit less pronounced effect in the distributions without a gap (solid lines).
 }\label{fig:ICs}
\end{figure}

Observations of young ($\sim$ few Myrs old) massive clusters suggest power law distributions for period and eccentricity (with different indices). The most conservative choice of initial conditions assumes that the final distributions remain unchanged from the birth distributions. The most agnostic choice of initial conditions assumes uniform distributions for period and eccentricity. Overall, we test six birth distribution combinations of eccentricity and period. Below, the abbreviation for each initial condition is given in parentheses. We use these abbreviations in the simulation labels.

We use three possible initial distributions for eccentricity. For populations of massive stars, observations indicate that the probability density distribution is either $f(e) \propto e^{-0.5}$ (``e05'') \citep[e.g.,][]{Sana+12,Sana&Evans2011} or uniform (``e1'') \citep[e.g.,][]{Alm2017}. We supplement these simulations with ones that assume a thermal eccentricity distribution (``e2'').

For the period distribution, we adopt a probability density of the form $f(\log {P}/{\textrm{days}}) \propto (\log {P}/{\textrm{days}})^{-0.5}$ (``PS'') \citep[e.g.,][]{Sana+12}. Since this observed probability density does not necessarily represent the birth distribution, we also use a uniform distribution in logarithmic space (``PU'') for certain simulations. Additionally, \citet{Sana2017} present observations of eleven young ($< 1 ~\mathrm{Myr}$) massive binaries in the open cluster M17 with a low radial-velocity dispersion. They offer multiple explanations for the peculiar velocity dispersion. In one of these explanations, the massive binaries form at larger periods, with some minimum birth separation, and later tighten and circularize. Specifically, \citet{Sana2017} suggest that a lower period cutoff of nine months is consistent with the velocity dispersion in M17 within $1\sigma$. To explore this intriguing possibility, we include a lower period cutoff of nine months in multiple (``gap'') simulations.

\subsection{Stability Criteria} 
To ensure that the secular approximation is valid, we apply three criteria to the initial conditions \citep{NF}.
The \citet{Mardling+01} criterion ensures the long-term stability of the system, where $i_{tot}$ denotes the mutual inclination between the inner and outer orbits in radians:
\begin{equation}
\label{eq:Mardling}
\frac{a_2}{a_1} > 2.8 \left( 1+\frac{m_3}{m_1+m_2}\right)^{2/5} \frac{(1+e_2)^{2/5}}{(1-e_2)^{6/5}} \left(1-0.3i_{tot}\right)
\end{equation}
We note that this stability criterion does not consider the system's lifetime \citep[see, for example,][]{Myllari+18} and therefore under-predicts the number of systems that may undergo EKL oscillations during the $10$~Myr integration time. 

The second criterion compares the amplitudes of the octupole and quadrupole terms to verify that the perturber is weak, i.e. that higher order terms have a negligible effect. This criterion stipulates that $\epsilon$, the ratio of the octupole and quadrupole amplitudes, remains small:
\begin{equation}
\epsilon = \frac{a_1}{a_2}\frac{e_2}{1-e_2^2} < 0.1 \ ,
\label{eq:epsilon}
\end{equation}

The last criterion requires that the inner orbit falls outside the Roche limit to avoid an early merger, before secular effects occur. We use the Roche limit as defined by \citet{Eggleton1983}:
\begin{equation}\label{eq:RL}
L_{R,ij} = 0.49 \frac{(m_i/m_j)^{2/3}}{0.6(m_i/m_j)^{2/3}+\ln (1+(m_i/m_j)^{1/3})} \ ,
\end{equation}
where $j=1,2$ and $R_i$ denotes stellar radius. All of our systems begin with $a_1(1-e_1)L_{R,ij}>R_i$.

The stability criteria affect the simulated birth distributions. In Figure \ref{fig:ICs}, this effect is most apparent in the thermal eccentricity distribution with no gap. The lack of a gap allows small values of $a_1$. Given the greater abundance of small semi-major axes, the Roche limit criterion limits the number of high eccentricity systems. Similarly, the requirement that $a_1$ be small compared to $a_2$ for a hierarchical triple, a condition reinforced by Eq. \ref{eq:Mardling} and \ref{eq:epsilon}, curtails the number of inner binaries at large periods (Figure \ref{fig:ICs}). Due to this effect, the PU and PS gap period distributions strongly resemble each other, and the no-gap PU and PS distributions converge at large periods.

\subsection{Stopping Conditions}

\begin{figure}
    \centering
        \includegraphics[width=.5\textwidth]{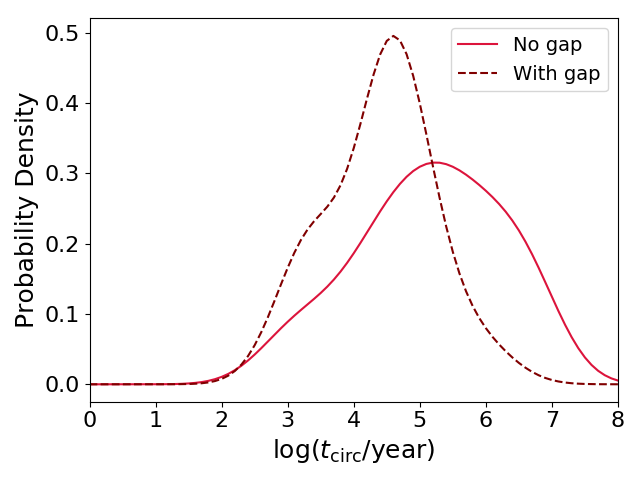}
    ~ 
    \caption{{\bf Circularization timescales} SET-PU-e1 (solid bright red curve) and SET-PS-e1-gap (dashed maroon curve) smoothed histograms of the circularization time ($t_{\mathrm{circ}}$), defined by our stopping condition, using a Guassian kernel density estimator. The no-gap simulation has an order of magnitude more circularized systems.}\label{fig:circtime}
\end{figure}
\begin{figure}
    \centering
        \includegraphics[width=.5\textwidth]{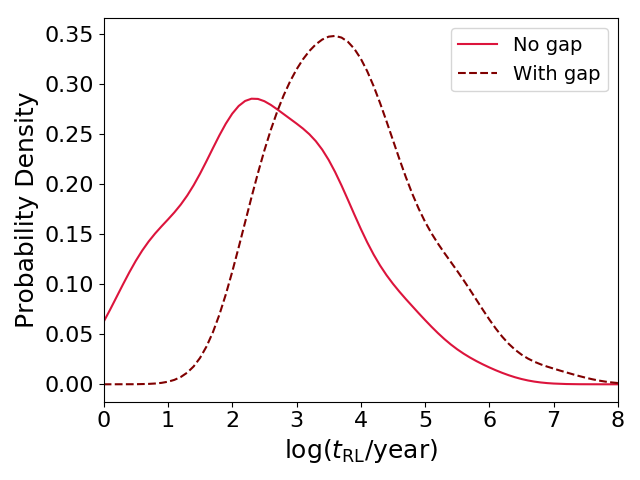}
    ~ 
    \caption{{\bf RL-crossing timescales} Smoothed histogram of the Roche Limit crossing time, called $t_{\mathrm{RL}}$, using a Guassian kernel density estimator for SET-PU-e1 (solid bright red curve) and SET-PS-e1-gap (dashed maroon curve). We term these systems RL binaries.}\label{fig:mergetime}
\end{figure}

We evolve each triple system for $10$~Myr. We also include conditions which, if met, result in an early termination of the integration. We consider two stopping conditions:

\begin{itemize}
    \item We terminate the simulation once the system tightens and circularizes because tide-dominated systems become numerically expensive. We consider a system to be tidally tightened and circularized when $a_1<2.1 R_{i,j}/L_{R,ij}$ and $e_1<0.001$. Figure \ref{fig:circtime} shows the typical circularization timescales for gap and no-gap systems.
    \item If an inner member have crosses the other's Roche limit, i.e., $a_1(1-e_1)L_{R,ij}<R_i$, see Eq.~\ref{eq:RL}, we terminate the integration. We denote the time upon which the system crosses the Roche limit as $t_{RL}$. In Figure \ref{fig:mergetime}, we show  typical $t_{RL}$. As expected, the gap systems have systematically longer Roche limit crossing times because they are associated with longer quadrupole timescales. The quadrupole timescale $T_{quad}$ depends on $P_{2}^2/P_1$, where $P_{1}$ ($P_2$) correspond to the inner (outer) orbital period. Typical gap systems have $\log {T_{quad}}>3.5$ with a median $\log T_{quad} \sim 6.7$. While the no-gap distribution has a similar median ($\log T_{quad} \sim 6.8$), the distribution is wider and has more systems with low quadrupole timescales. The lower limit of the distribution shifts to a shorter timescale, $\log {T_{quad}}>1.5$.
    
    Unlike \citet{NF}, we do not count these systems as merged products because we expect their merger times to be a few million years \citep{Stephan+16}, the same order as the stellar evolution timescale. The outcome of this interaction remains highly uncertain without comprehensive eccentric binary interaction physics. During this process, the stars may or may not be observed as two distinct stars. Thus, to differentiate these systems, we term them Roche limit-crossed binaries, or {\it RL} binaries for brevity.
    
    During a mass transfer process complicated by forced eccentricity oscillations from the tertiary, an observer may detect $m_1$ and $m_2$ as a binary system. We therefore define a possible observed orbit for the RL binaries with $ a_{\rm F} \sim  R_{\rm RL}$ by assuming angular momentum conservation, a plausible assumption during the final plunge of the merger. We also artificially set the eccentricities of RL binaries to $0.01$, which may inflate the number of circularized systems. We caution that the true properties of these systems are highly uncertain, and the actual observed periods may vary largely. In Figure \ref{fig:tides}, we denote this uncertainty with arrows.
   
\end{itemize}

\begin{figure*}
   \centering
        \includegraphics[width=\textwidth]{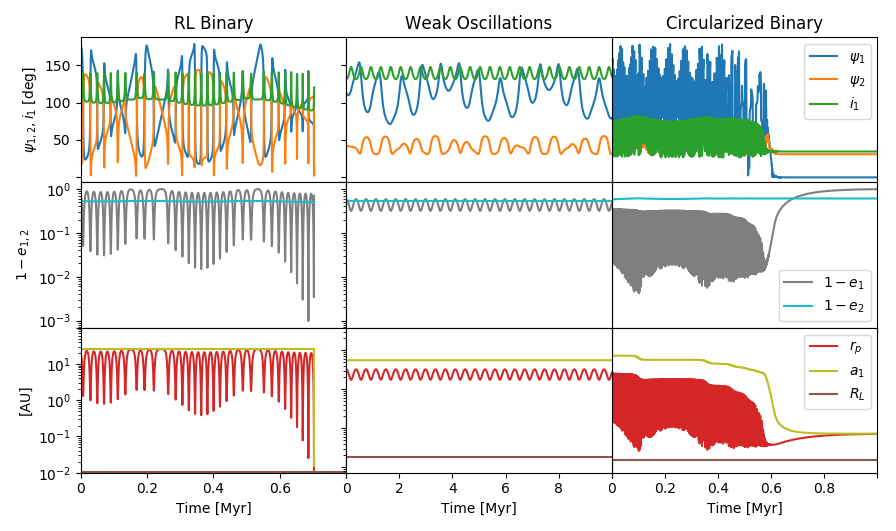}
    ~ 
    \caption{{\bf Archetypal systems} Here we show the time evolution of three triple systems. The left, middle, and right columns represent a Roche-limit crossing system, a system that undergoes weak oscillations, and a tidally circularized system, respectively. The first row shows the inner binary orbital inclination (blue) and obliquities (orange, green) of the stars; the second row, eccentricities of the inner and outer binary; and the last row, the inner semi-major axis, periapsis, and the Roche limit. Note that in the tidally tightened and circularized system, $r_p$ and $a_1$ converge while $e_1$ approaches zero. Both simulations take a $10~M_{\odot}$ primary. The left-hand side has initial values $m_2 = 2.1~M_{\odot}$, $m_3 = 11.7~M_{\odot}$, $a_1 = 25.8~\mathrm{AU}$, $a_2 = 286.7~\mathrm{AU}$, $e_1 = 0.22$, $e_2 = 0.47$, $\beta_1 = 178 ^{\circ}$, $\beta_2 = 33^{\circ}$, $i_1 = 104 ^{\circ}$, and $i_2 = 4^{\circ}$; the middle panel, $m_2 = 9.6~M_{\odot}$, $m_3 = 8.3~M_{\odot}$, $a_1 = 5.5~\mathrm{AU}$, $a_2 = 327.4~\mathrm{AU}$, $e_1 = 0.42$, $e_2 = 0.47$, $\beta_1 = 1.7^{\circ}$, $\beta_2 = 208^{\circ}$, $i_1 = 131^{\circ}$, and $i_2 = 4.07^{\circ}$; and the right-hand side, $m_2 = 6.0~M_{\odot}$, $m_3 = 12.6~M_{\odot}$, $a_1 = 7.2~\mathrm{AU}$, $a_2 = 85.8~\mathrm{AU}$, $e_1 = 0.77$, $e_2 = 0.42$, $\beta_1 = 44^{\circ}$, $\beta_2 = 49^{\circ}$, $i_1 = 71^{\circ}$, and $i_2 = 4^{\circ}$.} \label{fig:realtime}
\end{figure*}

\begin{figure}
    \centering
        \includegraphics[width=.5\textwidth]{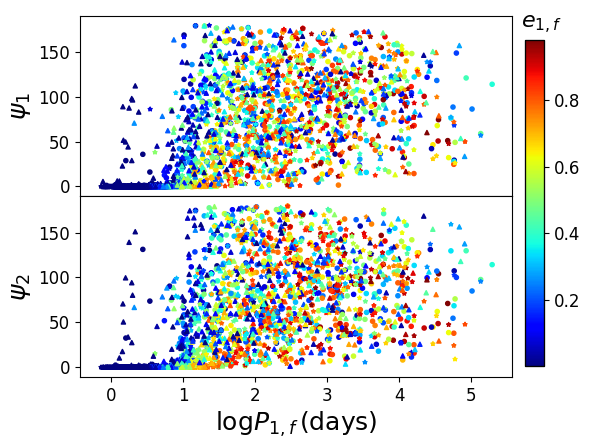}
    ~ 
    \caption{{\bf Spin-orbit angle} The final spin-orbit angle and period for both the primary (top) and secondary (bottom) colored by final eccentricity. Circles represent systems from MET-PU-e1, stars from MET-PU-e2, and triangles from MET-PS-e05. The points at low eccentricity and period but non-zero spin-orbit angles represent Cassini resonances. }\label{fig:beta}
\end{figure}
\section{Simulated Results} \label{Results}

\subsection{General outcomes}
Each system has three possible general outcomes, shown in Figure \ref{fig:realtime}. The final outcome depends mainly on the eccentricity excitations and the efficiency of the tides, which act to tighten and circularize the inner orbit. The ``strength'' of the EKL mechanism, parameterized by $\epsilon$ (Eq. \ref{eq:epsilon}), combined with the mutual inclination $i_{tot}$ of the system determine the nature of the eccentricity excitations \citep[e.g.,][]{Naoz16}.

\begin{itemize}
  \item {\bf Roche Limit (RL) crossing} When extreme eccentricity excitations occur on a shorter timescale than the tidal forces can circularize the system, the pericenter approach can become smaller than the Roche-limit (Eq.~\ref{eq:RL}). This type of behavior is depicted in the left hand side of Figure \ref{fig:realtime}. This outcome is common in the weak-tide regime examined in this paper.
    \item {\bf Tidal tightening and circularization} During periods of higher eccentricity induced by the third star, tidal forces can act to shrink and circularize the inner orbit. The tidal forces decrease the inner semi-major axis until the inner orbit decouples from the third star. The decoupling of the outer and inner orbits results in a conservation of their individual angular momenta. As a result, the typical final semi-major axis of the inner orbit is $\sim 2 a_{1,0}(1-e_{1,0})$, where the subscript ``0''  denotes initial values. The right column of Figure \ref{fig:realtime} shows an example of this evolution.
    
    Tidal dissipation also tends to align the spin axes of the stars with the inner orbit's angular momentum vector.  The spin-orbit angle $\psi_1$ ($\psi_2$) is defined as the angle between the inner orbit's angular momentum and the spin axis of $m_1$ ($m_2$). Figure \ref{fig:beta} shows that this angle goes to zero in most systems. However, in some cases, the system can become locked into a resonance, called a Cassini resonance:
    \begin{equation}\label{eq:spinorbit}
    \Omega_{j} = 2 \frac{2 \pi / P_1}{\cos{\phi_j}+\sec{\phi_j}} \ ,
    \end{equation}
    where $\phi_j$ denotes the spin-orbit angle and $\Omega_j$,  the spin period  of the two stars ($j=1,2$) in the inner orbit  \citep[e.g.,][]{Fabrycky+07,Stephan+16,NF}.
    \item {\bf Weak oscillations} In many cases, the third star induces only weak eccentricity and inclination oscillations that do not result in a dramatic change of the orbital parameters. The  middle column of Figure \ref{fig:realtime} shows an example system. While the locations of these individual systems in the eccentricity distribution shift slightly, they do not introduce a noticeable net change.
\end{itemize}

\begin{figure*}
    \centering
        \includegraphics[width=.94\textwidth]{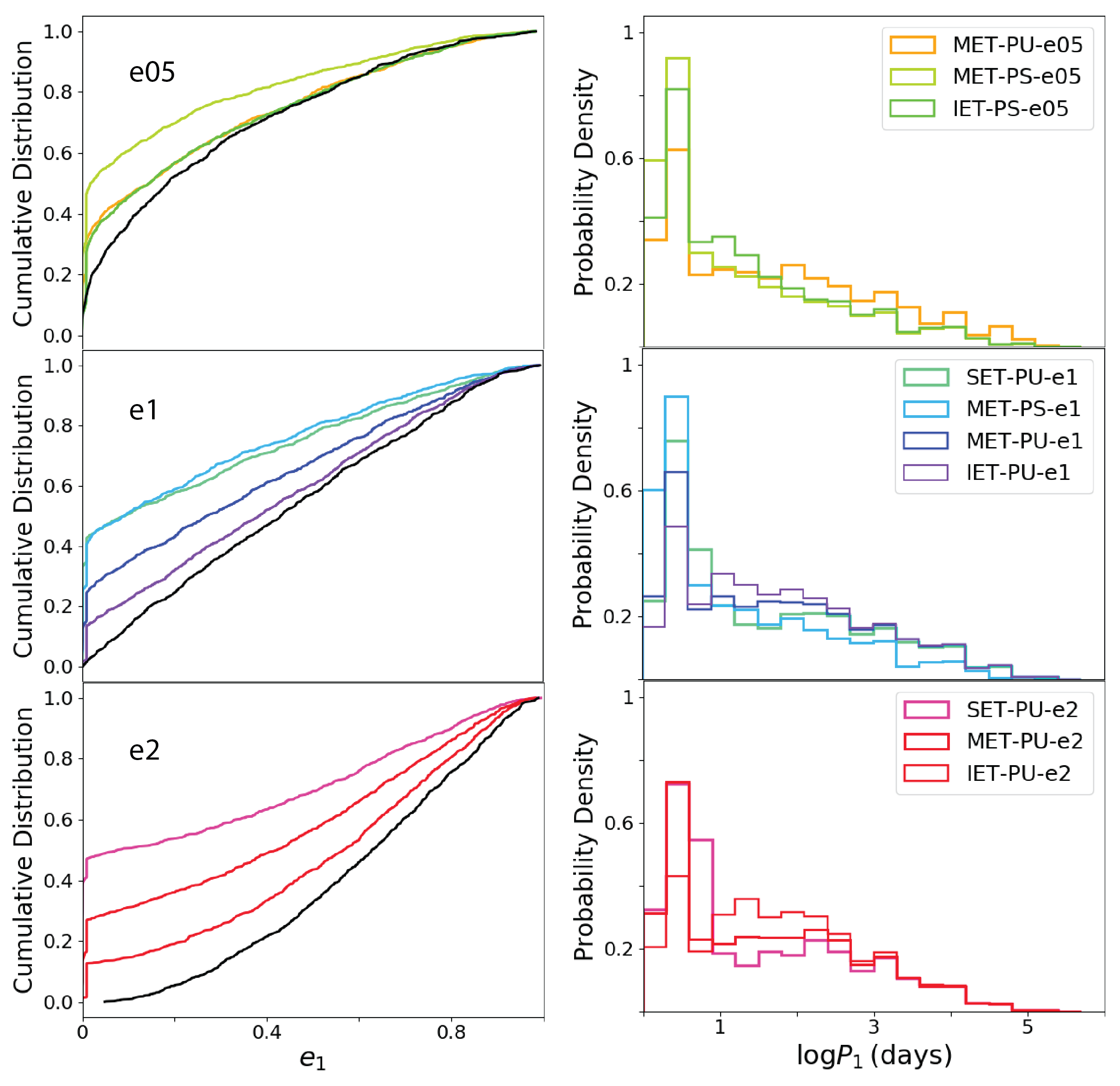}
    \caption{{\bf No-gap simulation results by eccentricity.} We consider the results for no-gap simulations by their eccentricity birth distribution. We create a color spectrum based on tidal efficiency and plot the birth eccentricity distribution in black. With the exception of low eccentricities where there is a build-up of circular systems, the final cumulative distribution strongly resembles the birth distribution.}\label{fig:Ecc_nogap}
\end{figure*}

\subsection{Eccentricity distribution and the effect of tides}
{\it The final eccentricity distribution tends to retain the same shape and curvature as the birth distribution irrespective of other conditions.} We quantify their similarity and provide predictions in \S \ref{sec:ePredic}. This result holds for gap and no-gap simulations. To visualize this trend, we group the results by initial eccentricity distribution in Figure \ref{fig:Ecc_nogap}. The initial period distribution has no discernible effect on the shape of the final eccentricity distribution.

As expected, tidal efficiency results in more circularized systems, increasing the vertical axis intercept of the cumulative distribution as shown in Figure \ref{fig:Ecc_nogap} for no-gap simulations. Less efficient tides result in an abundance of RL binaries, which can also increase the vertical axis intercept of the cumulative distribution. Recall that we artificially set eccentricity of the {\it RL} systems to $0.01$, which may inflate the number of circularized systems. Irrespective of efficiency, tides have no effect on the overall shape or functional form (e.g., thermal) of the distribution. 

\subsection{Period distribution and the effect of tides}
{\it The final long period distribution tends to retain the same shape and curvature  as the birth distribution irrespective of other conditions.} We quantify the similarity and provide predictions in \S \ref{period_traces}. 

\begin{figure}
    \centering
        \includegraphics[width=.5\textwidth]{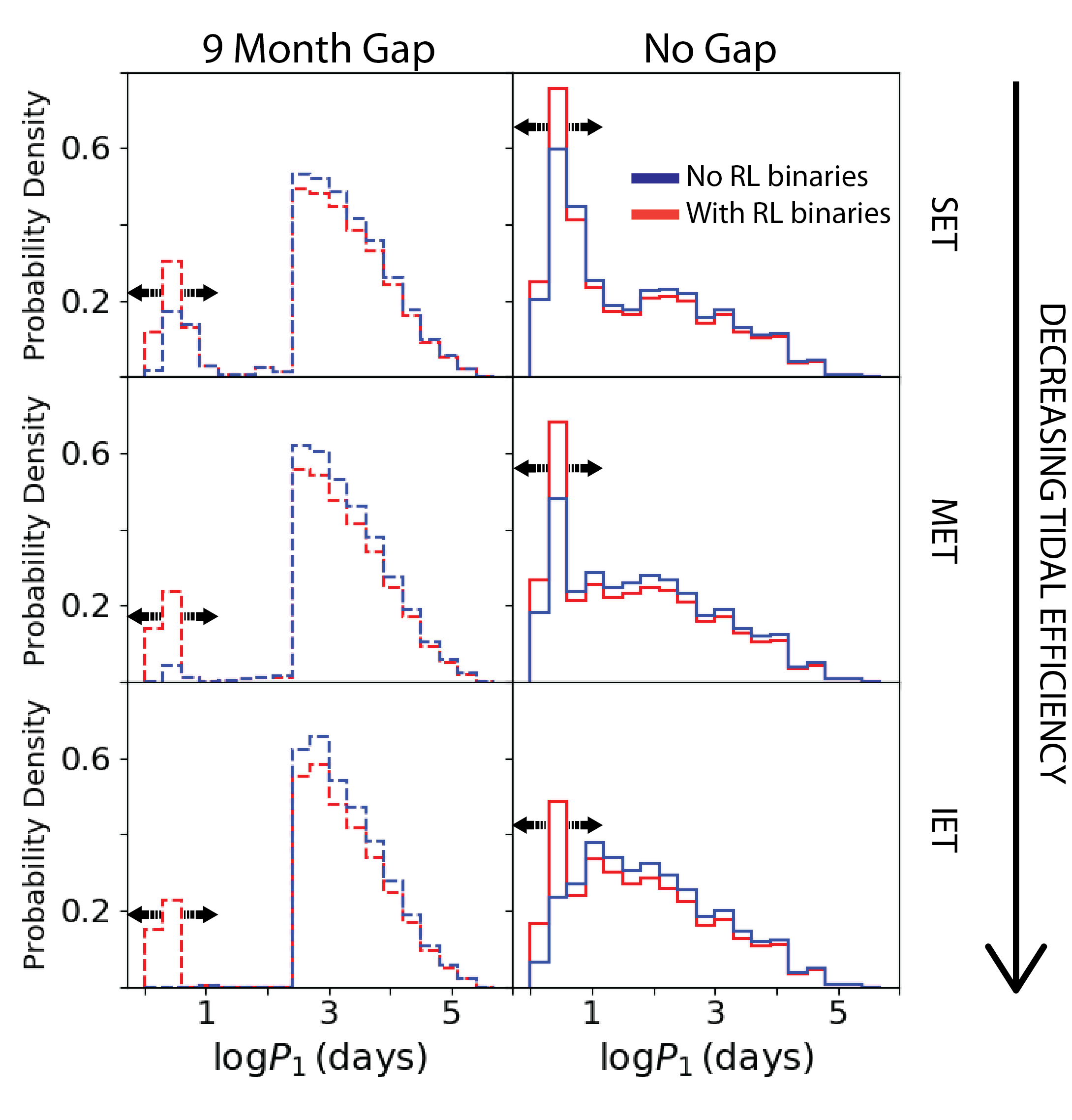}
    ~ 
    \caption{{\bf Tidal efficiency and final period distributions.} We show the final period distributions for three different tidal efficiencies: SET, MET, and IET, in order of greatest to least efficiency. The red distribution includes RL crossed systems as tight binaries with periods calculated from angular momentum conservation and the RL stopping condition. These systems therefore appear in a sharp, artificial peak. We use arrows to indicate that the width and true shape of this peak remain unknown. The blue distribution does not include RL crossed systems.}\label{fig:tides}
\end{figure}

As mentioned above, we treat the RL systems as tight binaries. Figure \ref{fig:tides} illustrates the strong dependency of the fraction of circularized systems on tidal efficiency. The dependency of RL system fraction on tidal efficiency is less pronounced, but generally less efficient tides result in more RL crossed systems. Figure \ref{fig:tides} exemplifies this trade-off between circularized systems and RL-crossed systems, where the red curves count RL binaries. An RL crossing occurs when the EKL mechanism drives extreme eccentricity excitations before before tides can shrink and circularize the system. A lower tidal efficiency therefore gives systems more chances to cross the RL. With the exception of the blue curves in Figure \ref{fig:tides}, we include RL binaries as short-period systems unless otherwise noted.

In Figure \ref{fig:tides}, the final period distribution for gap simulations appears bimodal and exhibits a dearth of systems at intermediate periods. The peak at short-periods consists of RL and tidally tightened systems. In tidally inefficient (IET) simulations, this peak comprises RL binaries, while the vast majority of systems remain at long periods.  Even with the most efficient tides (SET), $10$~Myr of EKL evolution fail to fill the gap.

\begin{figure}
 \includegraphics[width=0.45\textwidth]{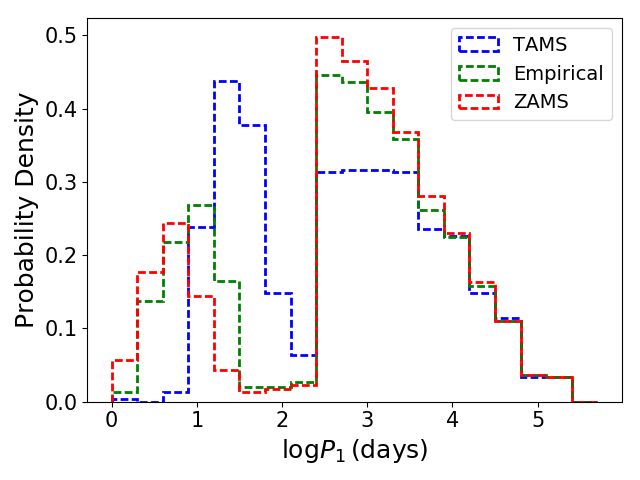}
 \caption{{\bf Stellar radii and final period distributions.} Here we show the dependence of the final inner binary period ($P_1$) probability density on the mass-radius relation for UET simulations with $t_v = 0.005$. The TAMS (red) and ZAMS (blue) relations yield the largest and smallest radius, respectively, for a given mass. Larger radii shift the peak to larger periods.}
 \label{fig:radii}
\end{figure}

Since tidal efficiency is highly sensitive to stellar radius, larger radii yield more circularized systems. Additionally, the final periods of these systems depend on the stellar radii due to angular momentum conservation \citep{FordRasio}. Therefore, as stars leave the main sequence and inflate, we expect circularized systems to have longer periods. To illustrate this behaviour, we consider three approaches to the mass-radius relation in Figure \ref{fig:radii}. Specifically, we consider the ZAMS mass-radius relation $R = 1.01 M^{0.57}$ (blue line), as well as $R = 1.33 M^{0.55}$ and the TAMS mass-radius relation, $R = 1.61 M^{0.81}$ \citep[e.g.,][]{Demircan} (green and red lines, respectively). To highlight the differences, we include a gap and make the tides unrealistically efficient (UET, unrealistic equilibrium tides). As shown in the figure, the larger radii correspond to longer periods for the tidally tightened binaries. Additionally, larger radii result in a more filled gap at intermediate periods. However, even with unrealistically efficient tides, the TAMS distribution fails to match observations.

\subsection{Predictions: Observable signatures of Birth Properties} \label{Signatures}

\subsubsection{Eccentricity}\label{sec:ePredic}

\begin{figure}
    \centering
        \includegraphics[width=.5\textwidth]{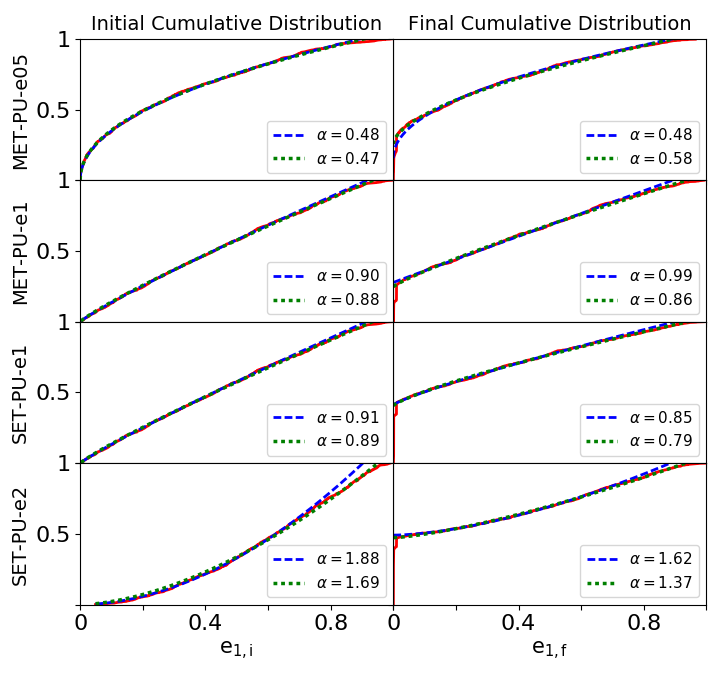}
    \caption{{\bf Observational signature of the initial eccentricity conditions.} We show the simulated cumulative distribution for select no-gap simulations in red for the final (initial) distribution, right (left) column.  
    Over-plotted is the fit $\mathrm{cdf}_e = \kappa_e  e_1^{\alpha_e} + \beta_e$ for the distribution calculated between two boundaries. In blue, dashed line, we mark the $0.1<e_1<0.7$ boundary while in green, dotted line, we mark the fit over the full range of eccentricity values. The fit function takes the following form:  $\mathrm{cdf}_e = \kappa_e e_1^{\alpha_e} + \beta_e$.
    The final distribution mirror the initial distribution and thus can serve as an observational signature.}\label{fig:fits}
\end{figure}

To quantify the degree of similarity between the simulated birth and final eccentricity distributions, we fit a function of the form 
\begin{equation}
    \mathrm{cdf}_e = \kappa_e e^{\alpha_e}+\beta_e \ ,
\end{equation}  
to the initial and final cumulative distributions, where $\alpha_e$ is the index of the power law and $\kappa_e$ and $\beta_e$ are constants. The parameters have the following limits: $0<\kappa_e<3$, $-2<\alpha_e<2$, and $0<\beta_e<0.6$. We argue that the vertical axis intercept, $\beta_e$, must be greater than or equal to zero because we are fitting a cumulative distribution. 
We fit both the initial and final distributions to show consistency between the two. As depicted in Figure \ref{fig:fits}, $\alpha_e$ changes by $0.3$ at most, and only for the most efficient tides. We caution that while this trend holds for hierarchical triple dynamics, we do not account for other dynamical processes that may alter the eccentricity distribution.

We vary the range of eccentricity values over which we fit. When fitting over the full range of eccentricity values, we impose the condition that $e>0.01$ to absorb all circularized and RL-crossed binary systems into the vertical-axis intercept. We also fit over the range $0.1 < e < 0.7$. We recommend fitting over this range because the EKL Mechanism tends to affect high eccentricity systems the most. As these more eccentric systems circularize or cross the Roche limit, they increase the number of low eccentricity systems. Similarly, a fit of the birth distribution over this range better reflects the function used to generate values because the stability criteria target the extremes.

\subsubsection{Orbital Period} \label{period_traces}

\begin{figure}
    \centering
        \includegraphics[width=.5\textwidth]{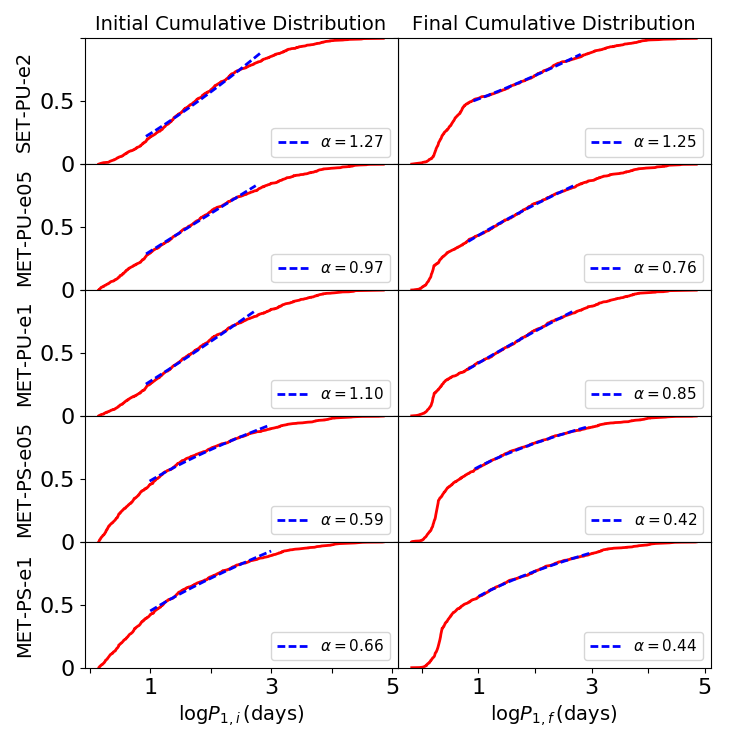}
    \caption{{\bf Traces of the initial orbital period distribution.} We show the simulated initial (left) and final (right) cumulative no-gap period distributions. We overplot a fit with of the form $\mathrm{cdf}_P = \kappa_P \left(\log P\right)^{\alpha_P} + \beta_P$. We fit the distribution over the range $1 <\log P < 3$.
    We find that a signature of the birth distribution is best preserved over this range.}\label{fig:fits_P}
\end{figure}

We use a function of the same form, $\mathrm{cdf}_P = \kappa_P \log P^{\alpha_P}+\beta_P$, with the same limits on $\alpha_P$, $\kappa_P$, and $\beta_P$. We fit the period distribution over the range $1<\log P < 3$. We select the lower limit to avoid the newly-formed peak of short-period systems. This lower limit must be adjusted to accommodate the width of the short-period peak, which will depend on the stellar radii (Figure \ref{fig:radii}) and therefore the age of the binary system when it circularizes. The reason for the upper limit is twofold: the stability criteria curb the number of long period systems, causing our initial distributions to converge at large period, and observational campaigns may not be sensitive to longer periods (e.g., \citet{Sana2013}). The birth distribution signature proves much more difficult to discern in the period distributions than in the eccentricity distributions. We find that the orbital period distribution is most consistently preserved over this range, $1<\log P < 3$. The power $\alpha$ changes by at most $\sim 0.3$ for PU simulations and stays $\leq 0.5$ for PS simulations.

\subsection{Comparison to observations} \label{Observations}

\subsubsection{Eccentricity Distribution}

Several groups have studied the eccentricity distributions of massive binaries.  Specifically, \citet{Alm2017} have examined OB-type spectroscopic systems in the Tarantula region to find that $40\%$ of systems have small eccentricities ($<0.1$). Furthermore, their eccentricity distribution appears uniform. However, the cumulative distribution flattens slightly for high ($>0.6$) eccentricities, indicating fewer systems there. In a study of $48$ massive systems from the Cygnus OB2 association, \citet{Kobulnicky2014} also show a flattening in the cumulative distribution at high eccentricity and attribute it to an observational bias. While this observed flattening may indeed reflect observational biases, we do find that our stability conditions curb the number of high eccentricity systems. As a result, our initial conditions exhibit flattening at large eccentricities (see Figure \ref{fig:ICs}), which persists in the final distribution. 

Similar to \citet{Alm2017}, \citet{Kobulnicky2014} note an abundance of low ($<0.1$) eccentricity systems and conclude that the distribution is uniform for $e \lesssim 0.6$. In their review, \citet{DK2013} also suggest that massive binaries follow a uniform eccentricity distribution, while \citet{Sana&Evans2011} and \citet{Sana+12} give a probability density of the form $f(e) \propto e^{-0.5}$. \citet{Moe} find a thermal eccentricity distribution, $f(e) \propto e$, for OB-type binaries using the catalog from \citet{Malkov+12}. However, their finding pertains to wider binaries with periods between 10 and 100 days, and many of our systems have shorter periods.

We find that the power law index is a good indicator of the initial condition. Additionally, \citet{Geller19} determine that star cluster dynamics has little to no effect on the shape of the eccentricity distribution for binaries with modest orbital periods. These combined results imply that the final distribution resembles the birth distribution. The birth distribution of population with uniform eccentricities is therefore also uniform.

\subsubsection{Period Distribution} \label{PeriodResults}

\citet{DK2013} estimate that $30 \%$ of massive binaries have periods of less than ten days. \citet{Sana&Evans2011} find that $50 \%$ to $60 \%$ of systems fall at $\log P < 1$. Similarly, the data of \citet{Alm2017} indicate an abundance ($\sim 40\%$) of short-period ($<1$ week) systems. \citet{Kobulnicky2014} also affirm the abundance of short-period binaries.

Counting RL binaries in the no-gap simulations, systems with $\log P < 1$ represent $45 \%$, $36 \%$, and $29 \%$ of the SET, MET, and IET inner binaries, respectively, in Figure \ref{fig:tides}. The super efficient tides (SET, first row) seem too efficient. However, counting only systems with $\log P < 3.4$ to reflect observational limits, the SET and MET simulations match the estimates of \citet{Sana&Evans2011} and \citet{Kobulnicky2014} with $51 \%$ and $42\%$ short-period ($<10$~days) systems, respectively. The IET simulation has $34\%$. Additionally, observations may not be sensitive to our so-called RL binaries. In that case, the IET tides become too inefficient to account for observations. Examining only tidally-tightened systems, the fractions of short-period binaries fall to $46\%$, $34\%$ and $24\%$ of total systems with $\log P < 3.4$ for SET, MET, and IET simulations, respectively.

A power law of the form $f(\log {P}/{\textrm{days}}) \propto (\log {P}/{\textrm{days}})^{\alpha}$ is often fitted to the orbital period distribution \citep[e.g.,][]{Kobulnicky2014}. The index of this power law varies in the literature. \citet{Sana+12} suggest $\alpha=-0.55$, while \citet{Sana2013} find $\alpha = -0.45$. \citet{Alm2017} conclude that an index between $-0.2$ and $-0.5$ reproduces the data well depending on the range over which they fit. \citet{Kobulnicky2014} suggest that tight binary peak has some structure, i.e. the distribution is not uniform at short-periods.

\begin{figure*}
    \centering
        \includegraphics[width=.8\textwidth]{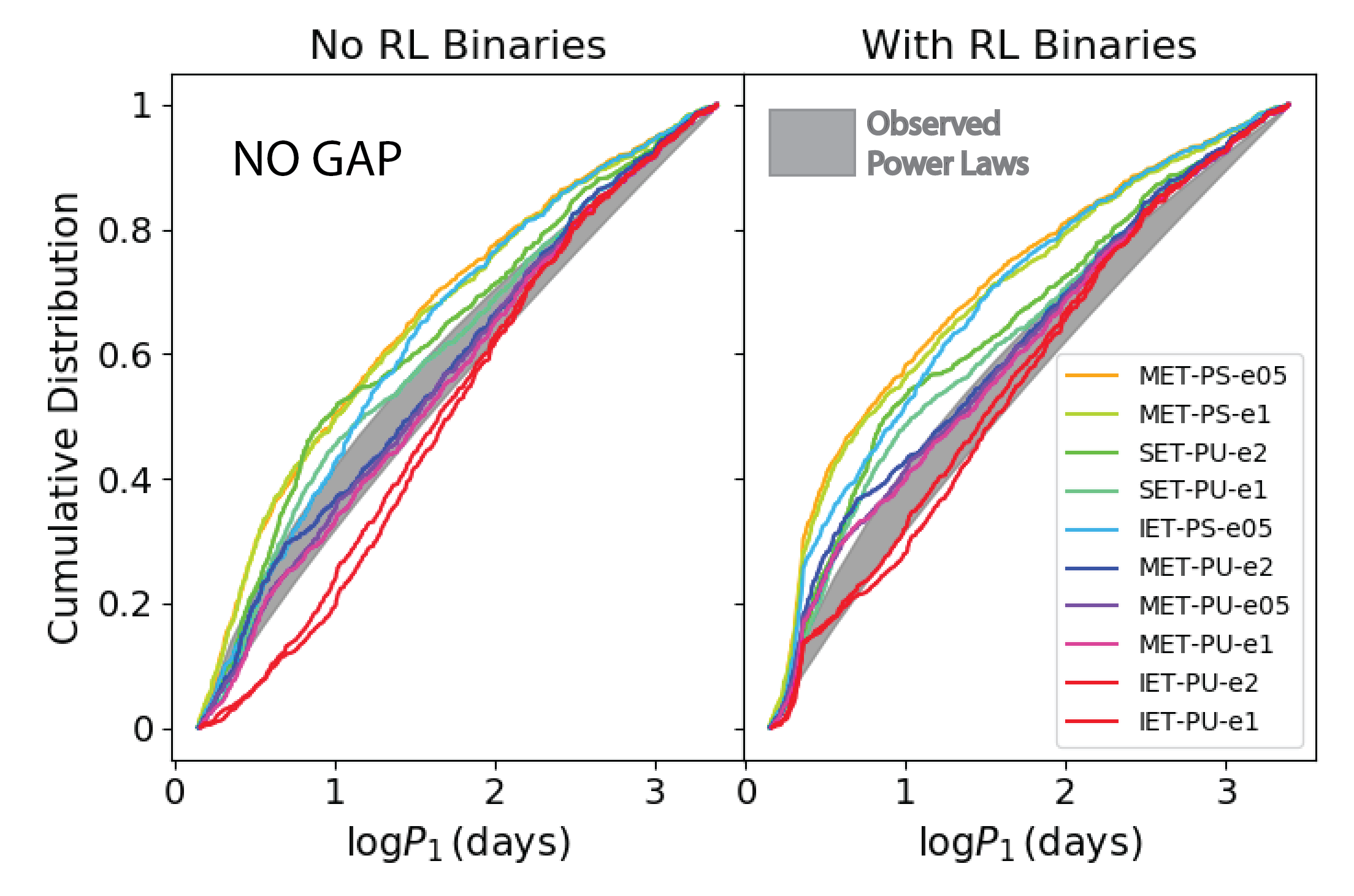}
    ~ 
    \caption{{\bf Power laws from observations compared with no-gap results} We plot the cumulative distributions for our no-gap simulations with and without RL binaries. The gray shaded region encompasses the range of power laws indicated by observations. The curves $\mathrm{cdf}_P \sim \left(\log P\right)^{0.8}$ \citep{Alm2017,Kobulnicky2014} and $\mathrm{cdf}_P \sim \left(\log P\right)^{0.5}$ \citep{Sana+12, Sana2013} bound the gray region. To avoid visual clutter, we deviate from our preferred presentation for the period distributions, a probability density, and instead focus on cumulative distributions here.}\label{fig:obs}
\end{figure*}

In Figure \ref{fig:obs}, we show the cumulative distributions for all of our no-gap simulations over the range $0.15<\log P < 3.4$. Bounded by the curves $\mathrm{cdf}_P \sim \left(\log P\right)^{0.8}$ \citep{Alm2017,Kobulnicky2014} and $\mathrm{cdf}_P \sim \left(\log P\right)^{0.5}$ \citep{Sana+12, Sana2013}, the gray region depicts the range of power laws fitted to observations. While several of our simulated results resemble the power laws fitted to the data, generally, SET and PS simulations seem to overpredict the number of short-period systems, while IET simulations underpredict the number of short period systems.

We perform a Kolmogorov-Smirnov (KS) two-sample test to compare our final period distributions to observations. We perform this test both with and without including RL binaries in our simulated sample. Comparing to the data of \citet{Sana+12}, we cannot reject the null hypothesis that the observations and simulated results share a parent probability density distribution for the following simulations: MET-PS-e05, MET-PS-e1, SET-PU-e2, IET-PS-e05 both with and without RL binaries. We also perform this test with the data from \citet{Kobulnicky2014} and find that we cannot reject the null hypothesis for MET-PS-e05 and IET-PS-e05 without RL binaries; MET-PU-e2 and MET-PU-e05 with RL binaries; and SET-PU-e2 and SET-PU-e1 both with and without RL binaries. We do not assign much significance to these results. We have a number of systems, the RL binaries, with poorly understood periods in MET and IET no-gap simulations, which hinder a statistical comparison between our results and the data. Additionally, the data have far fewer systems and are subject to observational biases: at large periods, the observed distribution likely diverges from the intrinsic distribution \citep{Sana+12}. The latter explains why the KS test favors PS and SET simulations, while the fitted power laws more often coincide with MET-PU simulations (Figure \ref{fig:obs}).

\begin{figure}
    \centering
        \includegraphics[width=.5\textwidth]{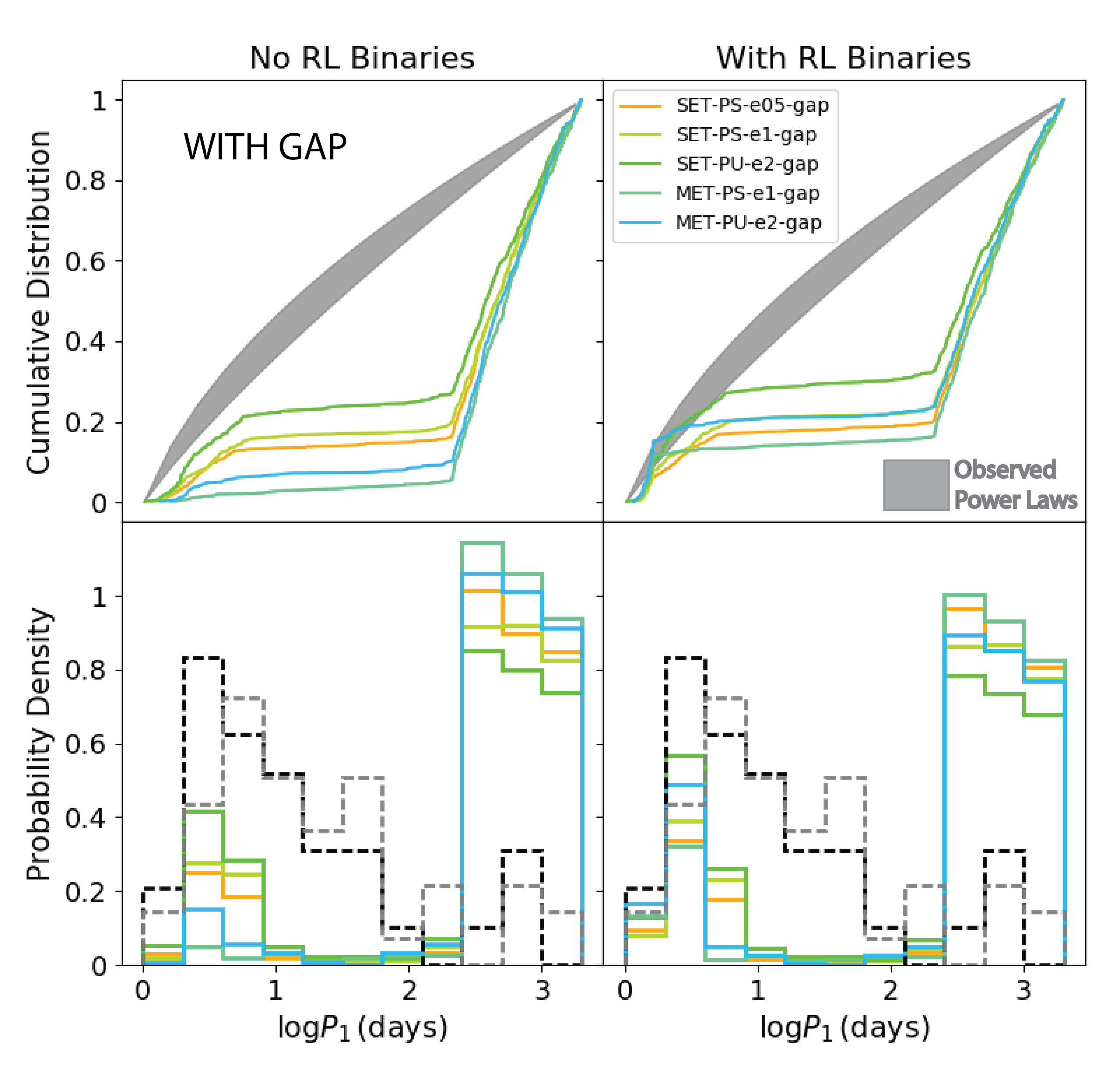}
    ~ 
    \caption{{\bf Upper Panel: Power laws from observations compared with gap results} We plot the cumulative distributions for select gap simulations with and without RL binaries. These simulations have more efficient (SET or MET) tides and therefore the most short-period systems. The gray shaded region represents observationally constrained power laws. The curves $\mathrm{cdf}_P \sim \left(\log P\right)^{0.8}$ \citep{Alm2017,Kobulnicky2014} and $\mathrm{cdf}_P \sim \left(\log P\right)^{0.5}$ \citep{Sana+12, Sana2013} bound this region. {\bf Lower Panel: Probability density with data} We plot the probability densities for select gap simulations with and without RL binaries. The gray and black dashed histograms represent data from \citet{Kobulnicky2014} and \citet{Sana+12}, respectively.
 }\label{fig:obs2}
\end{figure}

Unlike the no-gap simulations, the gap simulations seem to be in tension with observations. We plot the cumulative distributions for the gap simulations with more efficient tides in Figure \ref{fig:obs2}. The curves all exhibit a similar behavior indicative of a bimodal distribution. We again perform a KS test to compare our simulated results with the observations. The bottom panel of Figure \ref{fig:obs2} plots simulation results with the data from \citet{Sana+12} and \citet{Kobulnicky2014} that we test. We find that we can reject the null hypothesis that the simulated distributions and data are drawn from the same parent probability density distribution. Two major differences separate our simulated results from the observations: the persistence of a substantial population of long-period systems and the lack of intermediate period binaries, as illustrated by Figure \ref{fig:tides}. While the true separations of RL binaries -- and whether they can fall at intermediate periods -- is highly uncertain, a substantial long-period population will nonetheless persist.

\section{discussion}

A distant stellar companion can drive the long-term evolution of a massive stellar binary. Distant companions, which may be quite common, can therefore alter the observed orbital parameter distributions for massive binaries. We characterize these effects for a large variety of birth distributions and tidal efficiencies. We find the following: 

\begin{itemize}
  \item {\bf Spin-Orbit Angle Distribution:} Figure \ref{fig:beta} plots the spin-orbit angles and periods of 3000 realizations of massive triples color-coded by eccentricity. For long-period systems, no trend in spin-orbit angle and eccentricity exists. However, as systems circularize at periods of about 10 days, the orbit and stars' angular momenta align such that the spin-orbit angle goes to zero, with a few exceptions: occasionally the spin-orbit angle becomes locked in a (non-zero) resonance, called a Cassini resonance.
  
   \item {\bf Eccentricity distribution:} The final eccentricity distribution is an excellent indication of the birth distribution. The cumulative distribution retains the curvature of the birth distribution. Since the EKL Mechanism most affects high eccentricity systems, a fit over the range $0.1<e<0.7$ of a function $\mathrm{cdf_e} \propto e^{\alpha_e}$ quantifies this trend (see Figure \ref{fig:fits}). Generally, the change in $\alpha_e$ is within $0.3$, with a tendency to flatten -- or render more uniform -- the eccentricity distribution.
   \item {\bf Period distribution:} A signature of the birth period distribution persists at $1 < \log P < 3$. We fit the period cumulative distribution with a power law over this range. The index changes by $\sim 0.3$ for the uniform initial condition and remains $\lesssim 0.5$ for $\mathrm{cdf}_P \sim \left(\log P\right)^{0.5}$ (PS) initial condition simulations (Figure \ref{fig:fits_P}).
   
    \item {\bf Short period binaries:} Observations indicate an abundance of short-period binaries \citep[e.g.,][]{DK2013}. In our simulations, the tidal efficiency determines both the resulting number of short-period binaries and the dominant type of short-period system. Less efficient tides give systems more chances to cross the Roche limit, while efficient tides yield more circularized, tight binaries. The former results in an artificial peak in our final period distribution because the final properties of such systems remain uncertain.
    Due to angular momentum conservation, the final period of any short-period system is proportional to the stellar radius. We treat the stellar radii as constant. However, realistically, the radii will expand as the stars age. Systems which tidally tighten or cross the Roche Limit at later times will therefore fall at longer periods (Figure \ref{fig:radii}).
   
  \item {\bf Initial Period Gap:} \citet{Sana2017} suggest that massive binaries may form with large separations and tighten over time to match the parameter distributions of older populations. The EKL mechanism in concert with tidal dissipation represents a channel for producing hardened binaries. However, the EKL mechanism fails build up a sufficient population of short period binaries if we begin with a lower period cutoff of nine months (Figure \ref{fig:obs2}).
  
  We perform a brief calculation to assess whether type II migration may fill a 9 month gap in the inner period distribution. Following \citet{Armitage+07}, the type-II migration timescale $\tau$ can be written as: $\tau\sim {2}/({3 \alpha}) ({h}/{a_1})^{-2} \Omega^{-1}$, where $\alpha$ is related to the viscosity, $h$ is the scale height, and $h/a_1$ represents the disk aspect ratio. $\Omega$ denotes the angular velocity. A nine month period corresponds to a roughly 2 AU semi-major axis. Taking the primary and secondary to have masses $10$ and $5~M_{\odot}$, respectively, $\Omega$ is approximately $0.7~\mathrm{yr}^{-1}$. We assume that $h/a_1 \approx 0.07$ \citep[e.g.,][]{Ruge13} and $\alpha \approx 0.01$ \citep{Armitage+07}. We find that the timescale is about $2 \times 10^4~\mathrm{yr}$. Type II migration therefore represents a mechanism to bridge an initial period gap, as suggested by \citet{Sana2017}, with observations of older populations. Acting over a short timescale to tighten binaries, type II migration may fill the gap in the period distribution. If a binary system undergoing migration has a third companion, the gas will mostly suppress the gravitational perturbation of the tertiary star, although the system may develop an inclination between the disk and the companion \citep[e.g.,][]{Martin14}.

  \item {\bf Comparison with Observations: no initial period gap and moderate tides} We compare our cumulative distributions for the final inner orbital period with the observationally constrained power laws in Figure \ref{fig:obs}. Many of our results fall in the region bounded by the power laws indicated by the literature \citep[e.g.,][]{Alm2017}. Howerver, generally, simulations with moderately efficient tides (MET) and a uniform birth period distribution match observed distributions well.
  
  Several observations indicate a uniform eccentricity distribution except at high eccentricity \citep[e.g.,][]{Kobulnicky2014}. Only simulations that begin with a uniform eccentricity distribution produce a uniform distribution as the end result.
\end{itemize}

Unlike previous studies of EKL evolution in triple stellar systems \citep[e.g.,][]{NF,Bataille+18,MoeKratter18}, we find that the final period and eccentricity distributions carry a clear signature of the initial distributions. This behavior is a consequence of the short timescale of evolution ($\sim 10$~Myr) and the moderately efficient tides of stars with radiative envelopes. We expect that as stellar evolution increases the stellar radii and causes mass loss, the orbital configurations will significantly alter.

\section*{Acknowledgements}
SN thanks Howard and Astrid Preston for their generous support. SCR acknowledges support from the Eugene Cota-Robles Fellowship.

\bibliographystyle{mnras}
\bibliography{massivebinaries}







\appendix

\section{Simulation Parameters}\label{appendix:table}

We run several large Mont-Carlo simulations to cover a wide range of initial conditions. Table \ref{tab:10Myr} describes the 38 sets of Monte-Carlo simulations. Some simulations adopt a Kroupa mass function with limits $1 M_{\odot}<m<20 M_{\odot}$ (``K1'') or $6 M_{\odot}<m<20 M_{\odot}$ (``K6''). We find that the masses affect the results insofar as they determine the tidal efficiency. For example, for the former limits, the abundance of low mass stars with efficient tides will result in more circularized systems even when using the IET prescription. 

\begin{landscape}
\begin{table}
\begin{tabular}{|c|c|c|c|c|c|c|} 
\hline {\textbf{Label}} & {\textbf{Radius}} & {\textbf{Mass}} & {\textbf{Eccentricity}} & {\textbf{Period}} & {\textbf{Tides}} \\ \hline 
PU-e1-M10-TAMS & TAMS & $m_1=10 M_{\odot}$ \& Uniform $q$ & Uniform & Uniform & SET\\ 
\hline
PU-e1-M10-Emp & Empirical & $m_1=10 M_{\odot}$ \& Uniform $q$ & Uniform & Uniform & SET \\
\hline
PU-e1-M10-ZAMS & ZAMS & $m_1=10 M_{\odot}$ \& Uniform $q$ & Uniform & Uniform & SET, MET, IET, UET(g), SET(g), MET(g), IET(g) \\ 
\hline
PS-e1-M10-ZAMS & ZAMS & $m_1=10 M_{\odot}$ \& Uniform $q$ & Uniform & $f(\log P) \sim \log P^{-0.5}$ & SET, MET, SET(g) \\
\hline
PS-e05-M10-TAMS & TAMS & $m_1=10 M_{\odot}$ \& Uniform $q$ & $f(e) \sim e^{-0.5}$ & $f(\log P) \sim \log P^{-0.5}$ & UET(g) \\
\hline
PS-e05-M10-Emp & Empirical & $m_1=10 M_{\odot}$ \& Uniform $q$ & $f(e) \sim e^{-0.5}$ & $f(\log P) \sim \log P^{-0.5}$ & UET(g), LET(g) \\
\hline
PS-e05-M10-ZAMS & ZAMS & $m_1=10 M_{\odot}$ \& Uniform $q$ & $f(e) \sim e^{-0.5}$ & $f(\log P) \sim \log P^{-0.5}$ & MET, IET, UET(g), SET(g), MET(g) \\
\hline
PU-e05-M10-ZAMS & ZAMS & $m_1=10 M_{\odot}$ \& Uniform $q$ & $f(e) \sim e^{-0.5}$ & $f(\log P) \sim \log P^{-0.5}$ & MET \\
\hline
PU-e2-M10-ZAMS & ZAMS & $m_1=10 M_{\odot}$ \& Uniform $q$ & Thermal & Uniform & SET, MET, IET, SET(g), MET(g) \\
\hline
PS-e2-K6-Emp & Empirical & Kroupa with $6 M_{\odot} < m < 20 M_{\odot}$ & Thermal & $f(\log P) \sim \log P^{-0.5}$ & LET(g) \\
\hline
PS-e05-K6-Emp & Empirical & Kroupa with $6 M_{\odot} < m < 20 M_{\odot}$ & $f(e) \sim e^{-0.5}$ & $f(\log P) \sim \log P^{-0.5}$ & LET(g) \\
\hline
PU-e05-K6-Emp & Empirical & Kroupa with $6 M_{\odot} < m < 20 M_{\odot}$ & $f(e) \sim e^{-0.5}$ & Uniform & LET(g) \\
\hline
PU-e2-K6-Emp & Empirical & Kroupa with $6 M_{\odot} < m < 20 M_{\odot}$ & Thermal & Uniform & LET(g) \\
\hline
PS-e1-K1-ZAMS & ZAMS & Kroupa with $1 M_{\odot} < m < 20 M_{\odot}$ & Uniform & $f(\log P) \sim \log P^{-0.5}$ & IET(g) \\
\hline
PU-e1-K1-ZAMS & ZAMS & Kroupa with $1 M_{\odot} < m < 20 M_{\odot}$ & Uniform & $f(\log P) \sim \log P^{-0.5}$ & IET \\
\hline
PU-e1-K6-ZAMS & ZAMS & Kroupa with $6 M_{\odot} < m < 20 M_{\odot}$ & Uniform & $f(\log P) \sim \log P^{-0.5}$ & LET \\
\hline
PU-e2-K1-ZAMS & ZAMS & Kroupa with $1 M_{\odot} < m < 20 M_{\odot}$ & Thermal & Uniform & IET(g) \\
\hline
PS-e05-K1-ZAMS & ZAMS & Kroupa with $1 M_{\odot} < m < 20 M_{\odot}$ & $f(e) \sim e^{-0.5}$ & $f(\log P) \sim \log P^{-0.5}$ & IET(g) \\
\hline
PS-e2-K1-ZAMS & ZAMS & Kroupa with $1 M_{\odot} < m < 20 M_{\odot}$ & Thermal & $f(\log P) \sim \log P^{-0.5}$ & IET \\
\hline
PU-e05-K1-ZAMS & ZAMS & Kroupa with $1 M_{\odot} < m < 20 M_{\odot}$ & $f(e) \sim e^{-0.5}$ & Uniform & IET \\
\hline
PU-e05-K6-ZAMS & ZAMS & Kroupa with $6 M_{\odot} < m < 20 M_{\odot}$ & $f(e) \sim e^{-0.5}$ & Uniform & LET \\
\hline
\end{tabular}
\caption{Parameters and birth distributions used by each 10 Myr Monte Carlo simulation. A ``g'' denotes simulations that include a gap.}\label{tab:10Myr}
\end{table}
\end{landscape}

\section{Fitting the Period Distribution} \label{AdditionalFigures}

\begin{figure}
    \centering
        \includegraphics[width=0.5\textwidth]{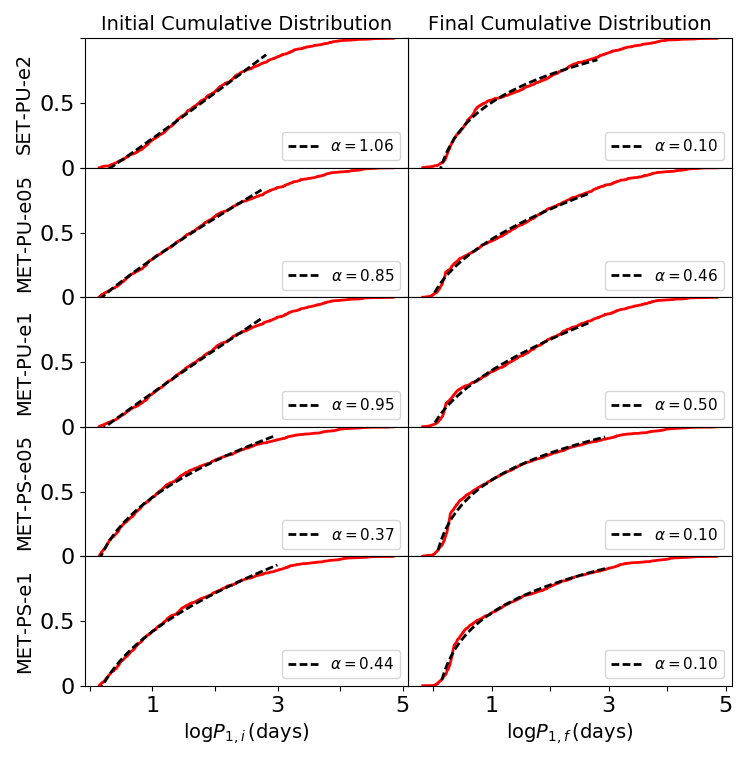}
    \caption{{\bf Final period distribution fit.} We show the simulated cumulative distribution in red for the final (initial) distribution, right (left) column. We overplot a fit with of the form $\mathrm{cdf}_P = \kappa_P \left(\log P\right)^{\alpha_P} + \beta_P$. We fit the distribution over the range $0.15<\log P < 3.4$ to examine overall trends in the period distribution.
    }\label{fig:fullfits_P}
\end{figure}

\begin{figure}
    \centering
        \includegraphics[width=.5\textwidth]{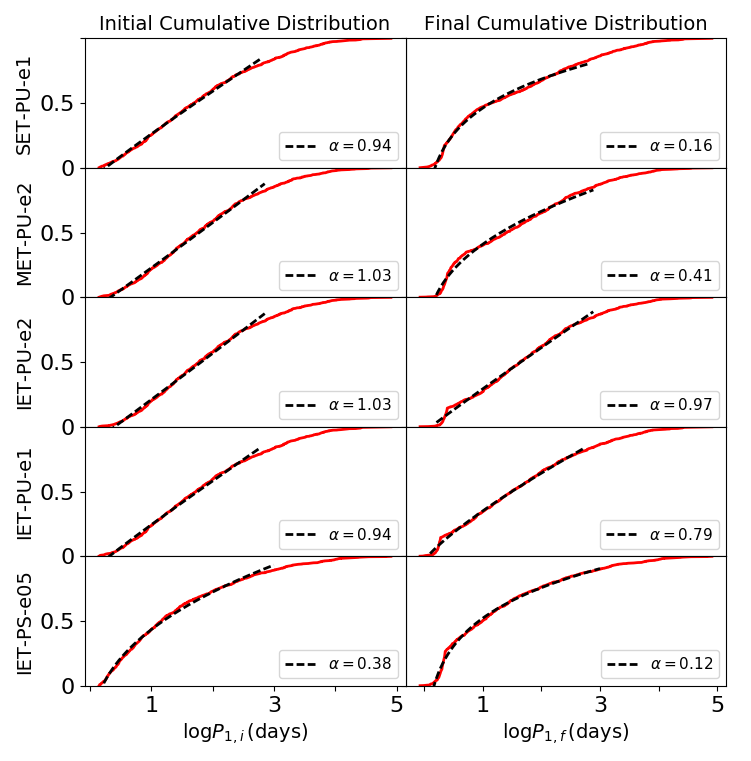}
    \caption{{\bf Final period distribution fit, additional examples.} We show the simulated cumulative distribution in red for the final (initial) distribution, right (left) column. We overplot a fit with of the form $\mathrm{cdf}_P = \kappa_P \left(\log P\right)^{\alpha_P} + \beta_P$. We fit the distribution over the range $0.15<\log P < 3.4$ to examine overall trends in the period distribution.
    }\label{fig:fullfits_P2}
\end{figure}

Taking a similar approach to Section \ref{period_traces}, we fit the cumulative no-gap period distribution over the range $0.15<\log P < 3.4$, selected to reflect the limits in, for example, \citet{Alm2017} and \citet{Sana2013}, in figure \ref{fig:fullfits_P}. With $0.4 < \alpha_P \leq 0.5$, PU-MET simulations best match observationally constrained cumulative distribution functions which have $0.5<\alpha<0.8$. We show additional examples in \ref{fig:fullfits_P2}. The IET-PU distributions are too flat, while either SET or PS conditions give too pronounced curvature with $\alpha_P < 0.2$. Taking the observations and our fits at face value, we suggest that moderately efficient tides -- or another model with a similar efficiency -- are the best candidate. However, these fits remain uncertain because of the RL binaries, included in the cumulative distribution.

\begin{figure}
    \centering
        \includegraphics[width=.5\textwidth]{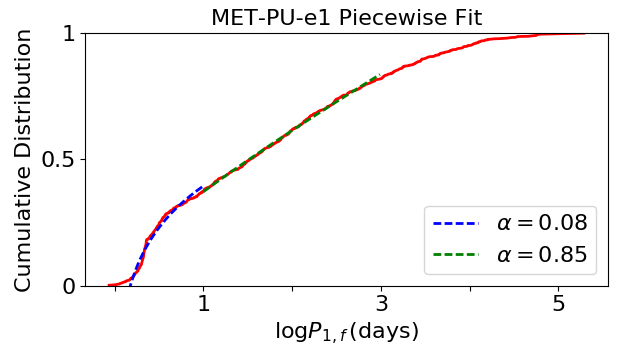}
    ~ 
    \caption{{\bf Example two-part fit of period distribution }We fit the most promising (PU, MET) simulated results with in two parts, $0.15 <\log P < 1$ and $1 <\log P < 3$. We use the e1 simulation and note that the eccentricity does not affect this result.}
 \label{fig:piecewise}
\end{figure}

We further suggest that cumulative distributions be fitted in two parts: one where the short-period peak occurs and one for periods longer that $\sim 10$ days. Since MET-PU simulations yield a similar final period distribution irrespective of eccentricity initial condition, we fit the MET-PU-e1 results as an example in Figure \ref{fig:piecewise}.

\section{Additional Examples of Period Distribution Traces} \label{more_examples}

We apply the method described in Section \ref{period_traces} to all of our no-gap simulations to find traces of the initial period distribution in the final result. We show the other five simulations here. As noted in Section \ref{period_traces}, the index $\alpha$ of the power law fitted over the range $1 <\log P < 3$ changes by at most $\sim 0.3$ for PU simulations and stays $\lesssim 0.5$ for PS simulations.

\begin{figure}
    \centering
        \includegraphics[width=.5\textwidth]{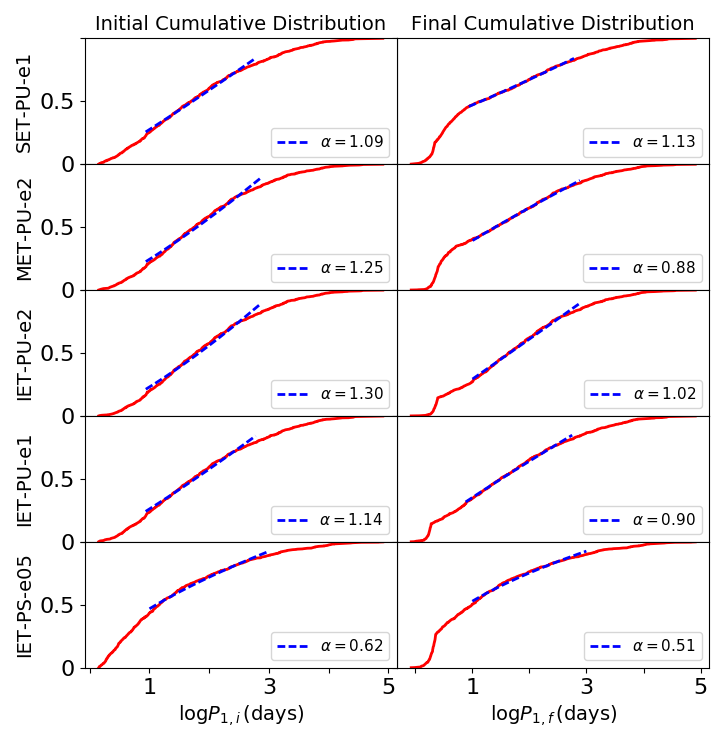}
    \caption{{\bf Signature of the initial orbital period distribution.} We show the simulated cumulative distribution in red for the final (initial) distribution, right (left) column. We overplot a fit with of the form $\mathrm{cdf}_P = \kappa_P \left(\log P\right)^{\alpha_P} + \beta_P$. We fitted the distribution over the range $1<\log P < 3$ to find a signature of the birth distribution.
    }\label{fig:fits_P2}
\end{figure}

\bsp	
\label{lastpage}
\end{document}